\documentclass[twocolumn,showpacs,amsfonts,aps,prc,floatfix,superscriptaddress,preprintnumbers]{revtex4-1}

\usepackage[colorlinks=true, pdfstartview=FitV, linkcolor=red, citecolor=blue, urlcolor=blue]{hyperref}
\usepackage{amsmath}
\usepackage{bm}
\usepackage{color}
\usepackage{graphicx}

\begin{document}

\title{Collinear parton splitting in early thermalization and chemical equilibration
}

\author{Akihiko Monnai}
\email[]{amonnai@riken.jp}
\affiliation{RIKEN BNL Research Center, Brookhaven National Laboratory, Upton, NY 11973, USA}

\author{Berndt M\"{u}ller}
\email[]{mueller@phy.duke.edu}
\affiliation{Department of Physics, Duke University, Durham, North Carolina 27708, USA}
\affiliation{Brookhaven National Laboratory, Upton, NY 11973, USA}

\date{\today}

\begin{abstract}
Early local equilibration of a hot medium created in high-energy heavy ion collisions has been one of the long standing issues in hadron physics. The glasma model predicts that the medium initially has a large amount of high-momentum gluons but few quarks. We develop a phenomenological model based on collinear splitting and recombination of quarks and gluons for a simplified description of thermalization and chemical equilibration. We find that both could be achieved in a short time but chemical equilibration is slower than thermalization.
\end{abstract}


\maketitle

\section{Introduction}
\label{sec1}
\vspace*{-2mm}

The quark-gluon plasma (QGP) \cite{Yagi:2005yb} exhibits near-perfect fluidity \cite{Kolb:2000fha} around the quark-hadron crossover in ``Little Bangs" at the Relativistic Heavy Ion Collider (RHIC) in Brookhaven National Laboratory and the Large Hadron Collider (LHC) in European Organization for Nuclear Research. The QCD system in high-energy heavy ion collisions is considered to go thorough several stages from color glass condensate (CGC) \cite{McLerran:1993ni} to glasma \cite{Lappi:2006fp}, QGP/hadronic fluid, and hadronic gas. In the past decades, there have been significant developments in the theoretical and numerical studies of relativistic viscous hydrodynamics \cite{Schenke:2010rr} and hadronic transport \cite{Bass:1998ca,Nara:1999dz} so that they now succeed in describing quantitatively the dynamics of the bulk medium after local equilibration. Investigation of the pre-equilibrated medium is expected to be a next frontier of the QCD collider physics.

Large azimuthal anisotropy in momentum space observed in the heavy ion experiments suggests the local equilibration of the medium occurs in less than $\sim 1$ fm/$c$. The early equilibration of the quark-gluon matter has been a topic of great interest especially because the glasma picture 
seems to exhibit large pressure anisotropy in the longitudinal direction partially due to rapid expansion of the system. 
So far various theoretical frameworks have been proposed to understand local equilibration of the hot and dense medium \cite{Geiger:1991nj, Wong:1996ta, Molnar, Baier:2000sb, Bass:2002fh, Krasnitz:2002mn, Lappi:2003bi, Arnold:2003rq, Wong:2004ik, Xu:2004mz, Arnold:2004ti, Rebhan:2004ur, Mueller:2005un, Mrowczynski:2005ki, Romatschke:2005pm,Janik:2006gp,Fujii:2008dd,Iwazaki:2008xi, Kunihiro:2008gv,Fujii:2009kb, Dusling:2010rm, Balasubramanian:2011ur, Kurkela:2011ti, Heller:2011ju, Fukushima:2011nq, Blaizot:2011xf, Berges:2011sb,Berges:2012us,Berges:2012ev, Dusling:2012ig, Schlichting:2012es, Attems:2012js,Berges:2012cj,Huang:2013lia,Berges:2013eia,Blaizot:2013lga, Fukushima:2013dma,Blaizot:2014jna}.

The CGC model indicates that the energy of colliding nuclei is mostly carried by gluons at high-energies and the gluon distribution is saturated at $\sim 1/\alpha_s$ up to the saturation momentum $Q_s$. This implies that the distribution has relatively large amount of high-momentum gluons compared to the thermal equilibrium distribution with the same momentum density. 
It is also important to note that a thermalized quark-gluon plasma should have large portion of quarks \cite{Gelis:2004jp} while the CGC has very few of them. Thus in addition to isotropization, local equilibration of a heavy ion system would require thermalization and chemical equilibration of quarks and gluons. 

In this work, we focus on describing thermal and chemical equilibration in a transverse direction and discuss collinear parton splitting for a source of quick production of low-momentum gluons and equilibrated quarks. Here we define thermalization in each direction and treat it as a separate issue from isotropization as the latter does not necessarily mean the former.
We introduce the recombination processes as required by the second law of thermodynamics and consider momentum smearing effects by Fokker-Planck-type diffusion for supplying off-shell partons. Numerical estimation is performed for qualitative understanding of the dynamical effects of collinear splitting and recombination on the parton distributions. 

Our model shares some philosophy with parton cascade \cite{Geiger:1991nj, Wong:1996ta, Molnar, Xu:2004mz}, bottom-up thermalization \cite{Baier:2000sb, Arnold:2003rq, Wong:2004ik, Mueller:2005un}, and recent Bose condensate approaches \cite{Huang:2013lia,Blaizot:2013lga} but differs from those methods in several aspects. Firstly, we aim to provide a simple and phenomenological framework of equilibration processes in a hot QCD matter and do not explicitly utilize Boltzmann equation. Secondly, recombination processes are introduced from the viewpoint of the second law of thermodynamics. Thirdly, we explicitly consider quarks and chemical equilibration of the system for describing the transition from gluon condensate to quark-gluon plasma. 

This paper is organized in the following manner. In Sec.~\ref{sec:2}, we model collinear splitting and recombination of quarks and gluons in the pre-thermalization stage. The off-shellness of partons are assumed to be provided by parton-medium interactions, which is implemented in the relativistic Fokker-Planck equation. We present numerical simulations of thermal and chemical equilibration from a color glass-like initial condition in one-dimensional non-expanding systems in Sec.~\ref{sec:3}. Sec.~\ref{sec:4} is devoted for summary and future prospects. We use the natural units $c = \hbar = k_B = 1$ in this paper.

\section{Collinear parton splitting model}
\label{sec:2}
\vspace*{-2mm}

We consider modification of quark and gluon phase-space distributions implied from collinear splitting and recombination processes. Effects of medium interaction, which induce smearing in momentum space, are taken into account via the drag and the diffusion effects of the relativistic extension of Fokker-Planck equation.

\subsection{Collinear parton splitting}
\begin{figure}[tb]
\begin{center}
\includegraphics[width=3.0in]{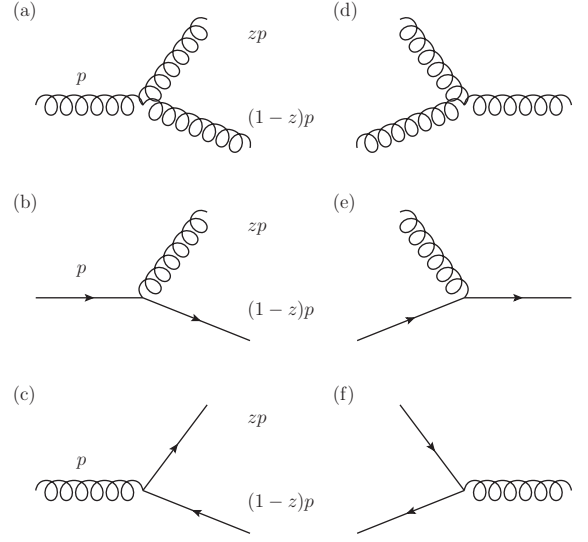}
\caption{\small Diagrams of (a) splitting of a gluon into two gluons, (b) gluon emission from a quark/antiquark, and (c) pair production of a quark and an antiquark from a gluon. Their reverse processes are shown in (d), (e), and (f), respectively.}
\label{fig:splitting}
\end{center}
\end{figure}

A parton phase-space distribution $f$ is modified when a splitting process occurs. Here we consider quarks and gluons for partons, \textit{i.e.}, $f = \{f_g, f_q\}$. We assume quarks and antiquarks are equal in number and include $q\bar{q}$ degrees of freedom in the quark distribution. Possible processes are (a) gluon splitting into two gluons $g \to g + g$, (b) quark/antiquark splitting into a gluon and a quark/antiquark $q/\bar{q} \to q/\bar{q} + g$, and (c) gluon splitting into a quark-antiquark pair $g \to q + \bar{q}$. The corresponding diagrams are shown in Figs.~\ref{fig:splitting} (a)-(c). One splitting of a parton with momentum $p_i$ into two with momentum fractions $zp_i$ and $(1-z)p_i$ where $0 \leq z\leq1$ alters the distribution as
\begin{eqnarray}
f(p_i) \to z^{-d} f\bigg(\frac{p_i}{z}\bigg) + (1-z)^{-d} f\bigg(\frac{p_i}{1-z}\bigg),
\end{eqnarray}
where $d$ is the number of dimensions. This satisfies the fact that the process should double the number density and conserve the momentum density:
\begin{eqnarray}
2 \int \frac{dp^d}{(2\pi)^d} f(p_i) &=& \int \frac{dp^d}{(2\pi)^d} \bigg[ z^{-d} f\bigg(\frac{p_i}{z}\bigg) \nonumber \\
&+& (1-z)^{-d} f\bigg(\frac{p_i}{1-z}\bigg) \bigg], \\
\int \frac{dp^d}{(2\pi)^d} p_j f(p_i) &=& \int \frac{dp^d}{(2\pi)^d} p_j \bigg[ z^{-d} f\bigg(\frac{p_i}{z}\bigg) \nonumber \\
&+& (1-z)^{-d} f\bigg(\frac{p_i}{1-z}\bigg)\bigg]. 
\end{eqnarray}
This implies that the time evolution of the phase space distribution by collinear splitting is phenomenologically expressed as
\begin{eqnarray}
\frac{\partial f_g(p_i)}{\partial t}|_\mathrm{sp} &=&\frac{1}{2} \int_0^{1} dz\ r_{gg}(z) [f_g^{z}(p_i)+f_g^{1-z}(p_i)-f_g(p_i)] \nonumber \\
&+& \int_0^{1} dz\ r_{gq}(z) f_q^{z}(p_i) - \int_0^{1} dz\ r_{qg}(z) f_g(p_i) \nonumber \label{eq:gluon_sp} \\
&\equiv& \mathcal{C}^g_\mathrm{sp}(p_i), 
\end{eqnarray}
\begin{eqnarray}
\frac{\partial f_q(p_i)}{\partial t}|_\mathrm{sp} &=& \int_0^{1} dz\ r_{gq}(z) [f_q^{1-z}(p_i) - f_q(p_i)]  \nonumber \\
&+& \int_0^{1} dz\ r_{qg}(z) [f_g^z(p_i)+f_g^{1-z}(p_i)] \nonumber \\
&\equiv& \mathcal{C}^q_\mathrm{sp}(p_i), \label{eq:quark_sp}
\end{eqnarray}
where the distributions of the split partons are expressed as $f^z (p_i) \equiv z^{-d} f(\frac{p_i}{z})$ and $f^{1-z} (p_i) \equiv (1-z)^{-d} f(\frac{p_i}{1-z})$. The factor $1/2$ is for identical particles. $r_{gg}(z)$, $r_{qg}(z)$, and $r_{gq}(z)$ determine the rates for the initial parton being split into two with the momentum fractions $zp_i$ and $(1-z)p_i$, which would be expressed with the splitting functions $R_{gg}$, $R_{gq}$, and $R_{qg}$ \cite{Altarelli:1977zs} as
\begin{eqnarray}
r_{gg}(z) &=& R_{gg}(z) \Gamma = 2 C_A \frac{[1 - z(1-z)]^2}{z(1-z)}  \Gamma, \\
r_{gq}(z) &=& R_{gq}(z) \Gamma = C_F \frac{1 + (1-z)^2}{z} \Gamma, \\
r_{qg}(z) &=& 2 N_f R_{qg}(z) \Gamma = 2 N_f T_R [z^2 + (1-z)^2] \Gamma,
\end{eqnarray}
where $C_A = N_c$, $C_F = (N_c^2 - 1)/N_c$, and $T_R = 1/2$. $N_f$ is the number of flavors and $N_c$ is that of colors. One has $2 C_A = 6$, $C_F = 4/3$, and $2 N_f T_R = 3$ when $N_c = 3$ and $N_f = 3$. $\Gamma$ is the emission rate of a parton. The splitting functions prefer the emission of soft gluons and have infrared divergences at $z=0$ and $1$ for $R_{gg}(z)$ and at $z=0$ for $R_{gq}(z)$. 

Here we make a rough estimation of the splitting rate $\Gamma$. It should be proportional to the coupling $\alpha_s$ as it is a one-vertex process. The transverse momentum $Q$ transferred from the medium while a parton travels the distance $L$ is given as
$Q^2 \sim \hat{q} L$ where $\hat{q}$ is the transport coefficient for momentum diffusion.
The rate should be dependent on $Q$ as it is the only scale; this implies $\Gamma \sim \alpha_s Q$ in the rest frame. In the Lorentz boosted frame this should be modified as $\Gamma \sim \alpha_s Q (Q/p) \sim \alpha_s \hat{q} L/p$ where $p = |\mathbf{p}|$. Since $L \sim 1/\Gamma$ for high-energy partons, we have
\begin{eqnarray}
\Gamma \sim \alpha_s^{1/2} \bigg( \frac{\hat{q}}{p} \bigg)^{1/2} . \label{eq:gamma}
\end{eqnarray}
In a thermal medium, one obtains $\Gamma_\mathrm{th} \sim \alpha_s^{3/2}T$ in terms of the temperature $T$ since $\hat{q} \sim \alpha_s^2 T^3$. It is important to note that the momentum transfer in a thermal medium $Q \sim \alpha_s^{1/4} T$ is larger than the effective mass $m_\mathrm{th} \sim \alpha_s^{1/2} T$ when the running coupling does not exceed unity, because this implies that the splitting is kinetically allowed. It should also be noted that the estimated coefficient reflects weakly-coupled nature of perturbation theory. 

\subsection{Parton recombination}
The splitting process should be suppressed when gluons become sufficiently dense because the second law of thermodynamics requires that the system is stable at local equilibrium. This can be phenomenologically implemented by introducing parton recombination terms in analogy to the Balitsky-Kovchegov (BK) equation \cite{Balitsky:1995ub}. Figs.~\ref{fig:splitting} (d)-(f) show the diagrams of such processes. Since the recombination rate should be determined by the densities of combining patrons rather than that of combined partons and completely cancel the splitting rate at equilibrium, one may express recombination as
\begin{eqnarray}
\frac{\partial f_g(p_i)}{\partial t}|_\mathrm{rc} &=& - \frac{1}{2} \int_0^1 dz\ \tilde{r}_{gg}(p_i,z) \nonumber \\
&\times& \frac{f_{g,\mathrm{eq}}^{z}(p_i)+f_{g,\mathrm{eq}}^{1-z}(p_i) - f_{g,\mathrm{eq}}(p_i)}{f_{g,\mathrm{eq}}(zp_i) f_{g,\mathrm{eq}}((1-z)p_i)} \nonumber \\
&\times& f_g(zp_i) f_g((1-z)p_i) \nonumber \\
&-& \int_0^1 dz\ \tilde{r}_{gq}(p_i,z) \frac{f_{q,\mathrm{eq}}^{z}(p_i)}{f_{g,\mathrm{eq}}(zp_i) f_{q,\mathrm{eq}}((1-z)p_i)} \nonumber \\
&\times& f_g(zp_i) f_q((1-z)p_i) \nonumber \\
&+& \int_0^1 dz\ \tilde{r}_{qg}(p_i,z) \frac{f_{g,\mathrm{eq}}(p_i)}{f_{q,\mathrm{eq}}(zp_i) f_{q,\mathrm{eq}}((1-z)p_i)} \nonumber \\
&\times& f_q(zp_i) f_q((1-z)p_i) \nonumber \\
&\equiv& \mathcal{C}^g_\mathrm{rc}(p_i), \label{eq:gluon_rc} 
\end{eqnarray}
\begin{eqnarray}
\frac{\partial f_q(p_i)}{\partial t}|_\mathrm{rc} &=& - \int_0^1 dz\ \tilde{r}_{gq}(p_i,z) \frac{f_{q,\mathrm{eq}}^{1-z}(p_i) - f_{q,\mathrm{eq}}(p_i)}{f_{g,\mathrm{eq}}(zp_i) f_{q,\mathrm{eq}}((1-z)p_i)} \nonumber \\
&\times& f_g(zp_i) f_q((1-z)p_i) \nonumber \\
&-& \int_0^1 dz\ \tilde{r}_{qg}(p_i,z) \frac{f_{g,\mathrm{eq}}^{z}(p_i) + f_{g,\mathrm{eq}}^{1-z}(p_i)}{f_{q,\mathrm{eq}}(zp_i) f_{q,\mathrm{eq}}((1-z)p_i)} \nonumber \\
&\times& f_q(zp_i) f_q((1-z)p_i) \nonumber \\
&\equiv& \mathcal{C}^q_\mathrm{rc}(p_i), \label{eq:quark_rc}
\end{eqnarray}
where $\tilde{r}(p,x)$ represents the rate for parton recombination constrained from momentum conservation and the requirement that it satisfies $\tilde{r}_\mathrm{eq}(p,x) = r(x)$ so that one has in equilibrium 
\begin{eqnarray}
\frac{\partial f_\mathrm{eq}(p_i)}{\partial t}|_\mathrm{sp} + \frac{\partial f_\mathrm{eq}(p_i)}{\partial t}|_\mathrm{rc} = 0,
\end{eqnarray}
for each parton component.

It should be noted that a na\"{i}ve relaxation equation, which only depends on the parton density at the momentum $p$, may exhibit unphysical behavior for the initial conditions with many partons in high momentum region because the recombination term can encourage them to form partons with even higher momentum while the lower momentum partons are still scarce. 

Finite density effects on the equilibrium distributions may be taken into account by introducing the effective thermal mass $m_\mathrm{th} \sim \alpha_s^{1/2} T$ which leads to in-medium equilibrium distributions
\begin{eqnarray}
f_{g,\mathrm{eq}}(p) &=& \frac{d_g}{\exp (\sqrt{p^2 + m_\mathrm{th}^2}/T) - 1} , \\
f_{q,\mathrm{eq}}(p) &=& \frac{d_q}{\exp (\sqrt{p^2 + m_\mathrm{th}^2}/T) + 1},
\end{eqnarray}
where $d_g$ and $d_q$ are the gluon and the quark degeneracies. $d_g = 16$ and $d_q = 36$ when $N_f = 3$. Both gluon and quark distributions are finite at $p=0$. $T$ is the temperature of a thermalized medium, which can be fixed from
the integrated momentum density
\begin{eqnarray}
P_i = \int \frac{dp^d}{(2\pi)^d} p_i [f_{g,\mathrm{eq}}(p) + f_{q,\mathrm{eq}}(p)].
\end{eqnarray}
The numbers of quarks and gluons at equilibrium for a given temperature are also determined. The entropy density of the system can be defined as
\begin{eqnarray}
s &=& d_g \int \frac{dp^d}{(2\pi)^d} [\tilde{f_g} \ln \tilde{f_g} - (1+\tilde{f_g}) \ln (1+\tilde{f_g})] \nonumber \\
&+& d_q \int \frac{dp^d}{(2\pi)^d} [\tilde{f_q} \ln \tilde{f_q} + (1-\tilde{f_q}) \ln (1-\tilde{f_q})],
\end{eqnarray}
where one-particle distributions are defined as $\tilde{f_g} = f_g/d_g$ and $\tilde{f_q} = f_q/d_q$. 

It should be noted that despite their similarities, our equations of motion of the phase-space distributions (\ref{eq:gluon_sp}), (\ref{eq:quark_sp}), (\ref{eq:gluon_rc}), and (\ref{eq:quark_rc}) are not Boltzmann equations. Also our approach is different from the BK equation since the distribution can be denser than $f  \sim 1/\alpha_s$ near $p=0$.

\subsection{Off-shell conditions}

The splitting processes are allowed when a parton interacts with the medium and becomes off-shell. The smearing effect in momentum space can be embedded in the model by relativistic Fokker-Planck equation, which takes account of parton-medium interactions. The equation reads for each component as
\begin{eqnarray}
\frac{\partial f}{\partial t}|_\mathrm{FP} &=& \frac{\partial}{\partial p_i} \bigg[ A^i f + \frac{\partial}{\partial p_j} (B^{i j} f) \bigg] 
= \mathcal{C}_\mathrm{FP},
\end{eqnarray}
where $f = \{ f_g, f_q \}$. The coefficients $A^i \sim \nu p^i$ and $B^{ij} \sim D \delta^{ij}$ characterize the drag and the diffusion terms in the small momentum limit. The equation is often employed for describing non-equilibrated heavy quarks in a medium \cite{Svetitsky:1987gq}. We apply it to the pre-equilibrated partons here. It should be noted that diffusion would contribute to isotropization, though it is not the main focus of this paper. The coefficients are related to one another in relativistic Einstein relation as $D = \nu E T$. 

We would like to roughly estimate the drag and the diffusion coefficients.
The pure diffusion equation is analytically known to yield a Gaussian distribution as the solution for a delta function initial condition. The standard deviation is related to the diffusion coefficient $D$ as $\sigma = \sqrt{2Dt}$. Characteristic time is that of the inverse splitting rate $1/\Gamma$ defined in Eq.~(\ref{eq:gamma}).
On the other hand, $\sigma$ is related to the momentum smearing of a parton after medium interaction. The transverse momentum given from the medium before a parton splitting modifies the overall momentum as $p+\Delta p \sim p (1 + Q^2/p^2)^{1/2} \sim p + Q^2/2p$. 
This implies 
\begin{eqnarray}
\sigma \sim \frac{Q^2}{2p} \sim \frac{\hat{q}}{2 p \Gamma}, 
\end{eqnarray}
and $\sigma_\mathrm{th} \sim \alpha_s^{1/2} T/2$ in thermal media for $p \gg Q$. Identifying the diffusion width at $t \sim 1/\Gamma$ with that of thermal momentum smearing, the drag and the diffusion coefficients in thermal media could be estimated as $\nu_\mathrm{th} \sim \alpha_s^{5/2} T$ and $D_\mathrm{th} \sim \alpha_s^{5/2} T^3$.

\section{Numerical Analyses}
\label{sec:3}
\vspace*{-2mm}

We investigate the equilibration processes of the parton splitting model qualitatively by numerical simulation. The time evolutions of the parton distributions are considered in one-dimensional non-expanding systems. We consider a CGC-like initial condition for the gluon distribution where $f_g(p < Q_s) \sim 1/\alpha_s$ and $f_g(p>Q_s) \sim 0$. We set $\alpha_s = 0.2$ and $Q_s = 2$ GeV. The initial quark distribution is set to vanishing, \textit{i.e.}, $f_q (p) = 0$ reflecting the fact that the pre-collision state is dominated by saturated gluons. The initial time would be of the order of the decoherence time $t \sim t_\mathrm{dec} \sim \mathcal{O}(1/Q_s)$ \cite{Muller:2005yu}. Here we choose the initial time $t = 0.2$ fm/$c$. 
The splitting ratio $\Gamma = \alpha_s^{3/2} T_\mathrm{eff}$ and the drag coefficient $\nu = \alpha_s^{5/2} T_\mathrm{eff}$ are used as model parameters according to the discussion in Sec.~\ref{sec:2} for demonstration. Here $T_\mathrm{eff}$ is defined as the temperature of the medium when it is locally equilibrated. The diffusion coefficient is given by the relation $D = \nu E T_\mathrm{eff}$. Infrared divergence, which exists in the splitting functions, can be dealt with by introducing a cut-off in momentum space.

\subsection{Gluon system}

We first consider a pure gauge system where only gluon splitting and recombination processes are present ($N_f = 0$). Fokker-Planck drag and diffusion effects are taken into account. In this case, thermalization is the dynamics of interest as chemical equilibration does not exist in the system a priori. The temperature for the equilibrium distribution $T = 0.711$ GeV is set so the momentum densities of the initial and the equilibrium distributions match. 

\begin{figure}[tb]
\begin{center}
\includegraphics[width=3.0in]{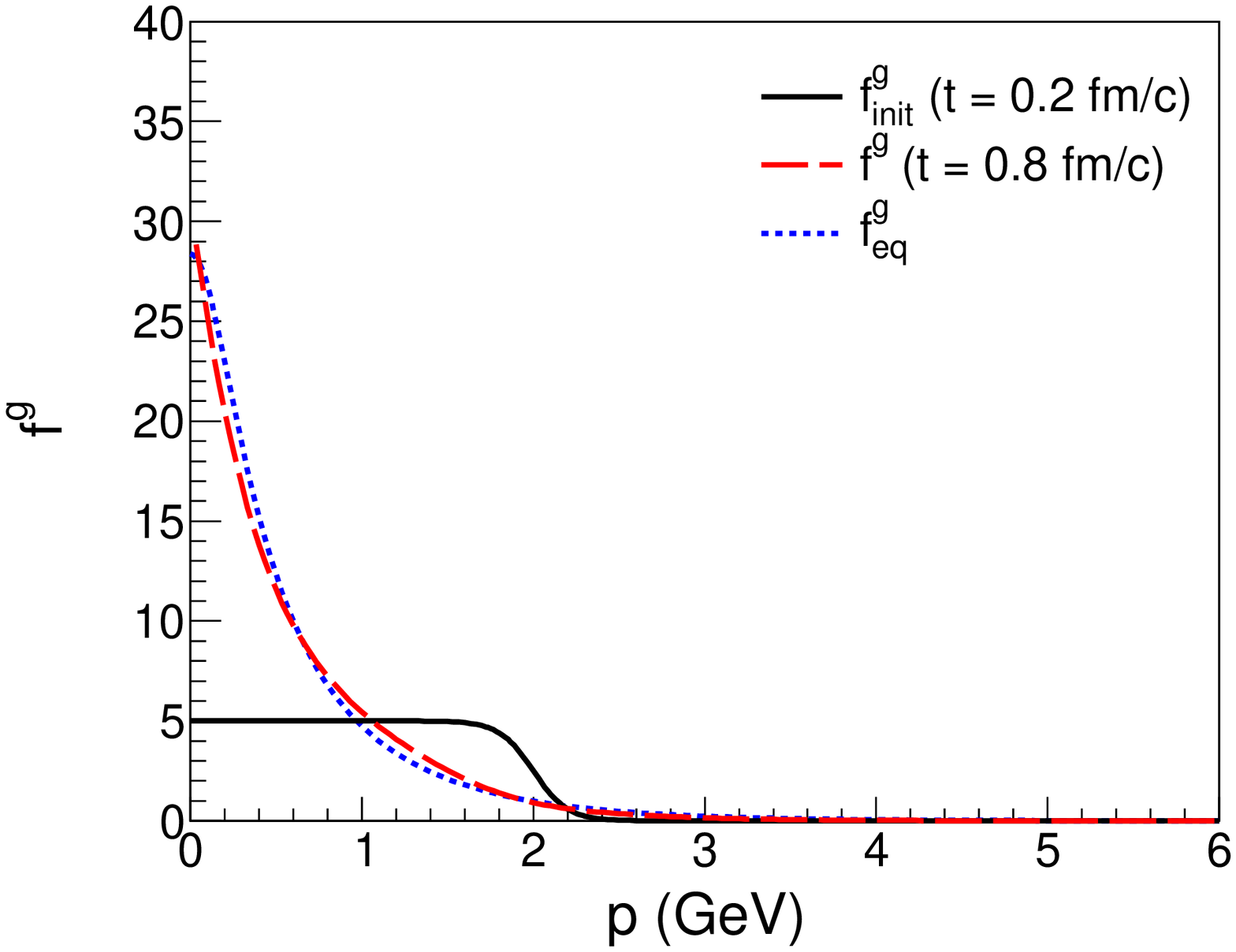}
\includegraphics[width=3.0in]{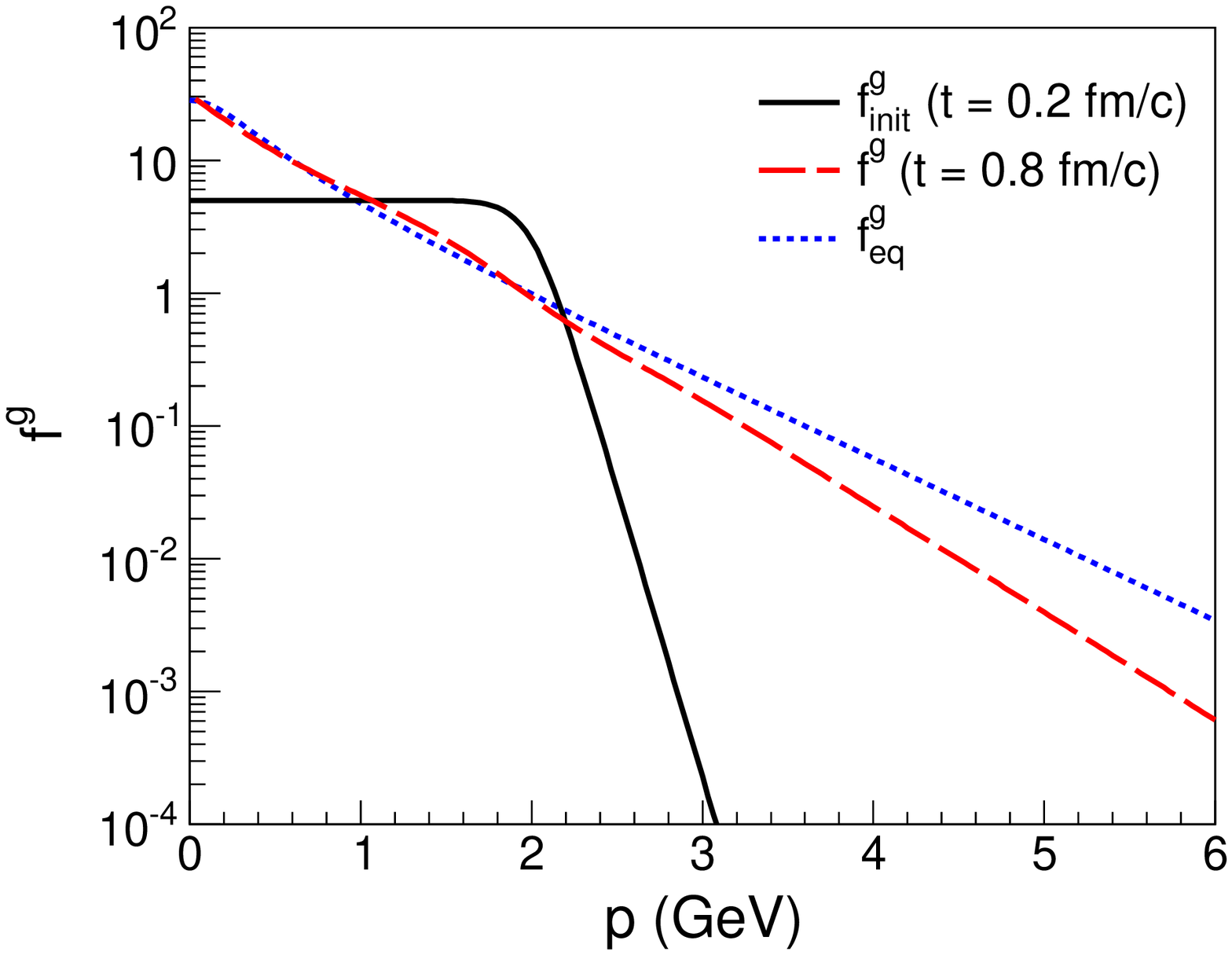}
\caption{\small (Color online) (a) The gluon phase-space distribution in the initial state ($t=0.2$ fm, solid line) and during the time evolution in a pure gauge system ($t=0.8$ fm, dashed line) compared with that in equilibrium (dotted line). (b) The same figure on a logarithmic scale.
}
\label{fig:f}
\end{center}
\end{figure}

The gluon distribution at $t = 0.8$ fm/$c$ is shown in Fig.~\ref{fig:f}. One can see that the splitting process introduces low-momentum gluons to the system and makes visible contribution to thermalization. The recombination process reduces infrared behavior at vanishing momentum and produces partons beyond the saturation momentum as one can see in the figure on a logarithmic scale. Near-complete transition to the equilibrium distribution can be found in the current parameter settings, suggesting that the splitting and recombination processes would play an important role in heavy ion dynamics. 
The entropy density is found to increase monotonously during the time evolution, satisfying the second law of thermodynamics.

\subsection{Quark-gluon system}

\begin{figure}[tb]
\begin{center}
\includegraphics[width=3.0in]{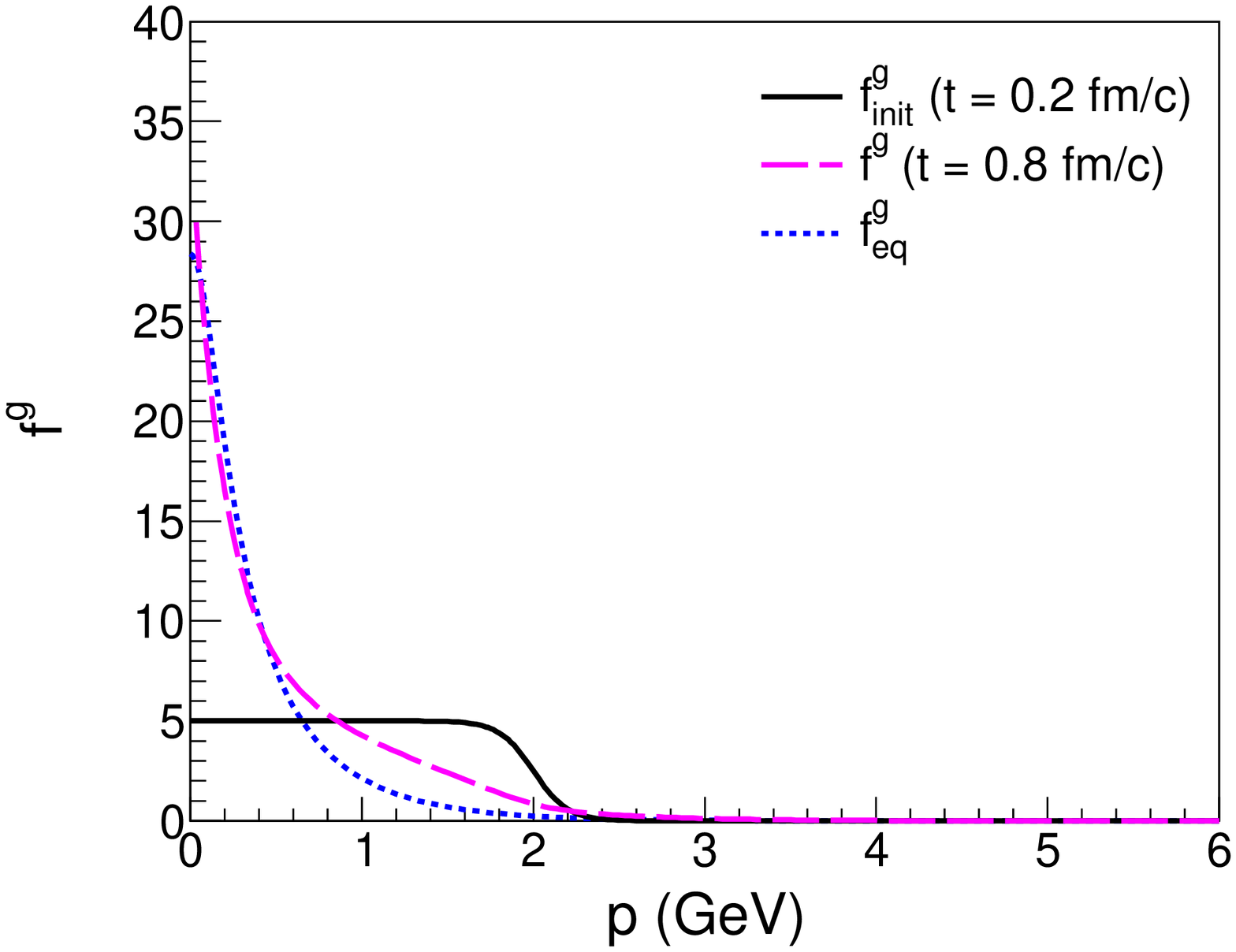}
\includegraphics[width=3.0in]{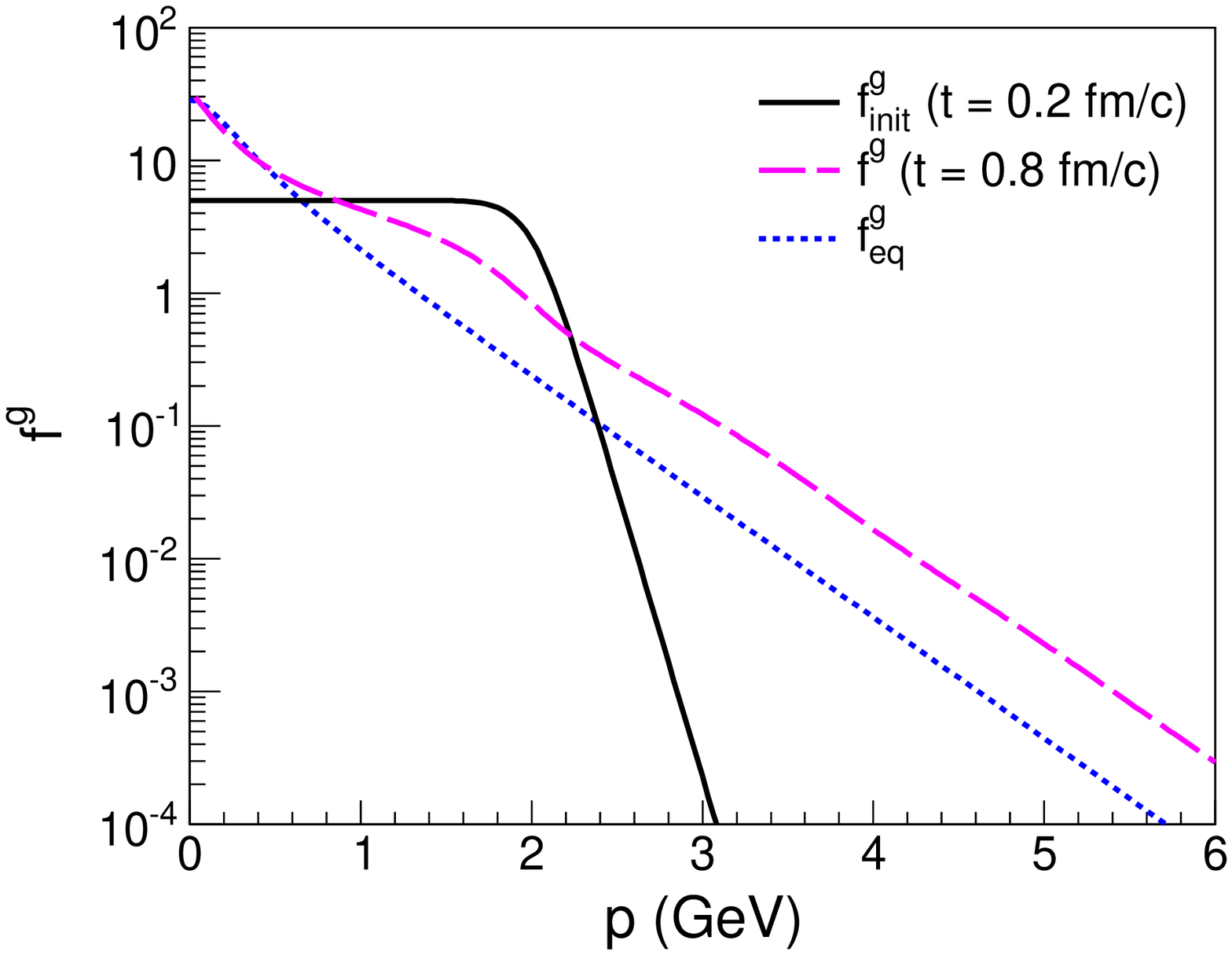}
\caption{\small (Color online) (a) The gluon phase-space distribution in the initial state ($t=0.2$ fm, solid line) and during the time evolution with $N_f = 3$ ($t=0.8$ fm, dashed line) compared with that in equilibrium (dotted line). (b) The same figure on a logarithmic scale.
}
\label{fig:fg}
\end{center}
\end{figure}

\begin{figure}[tb]
\begin{center}
\includegraphics[width=3.0in]{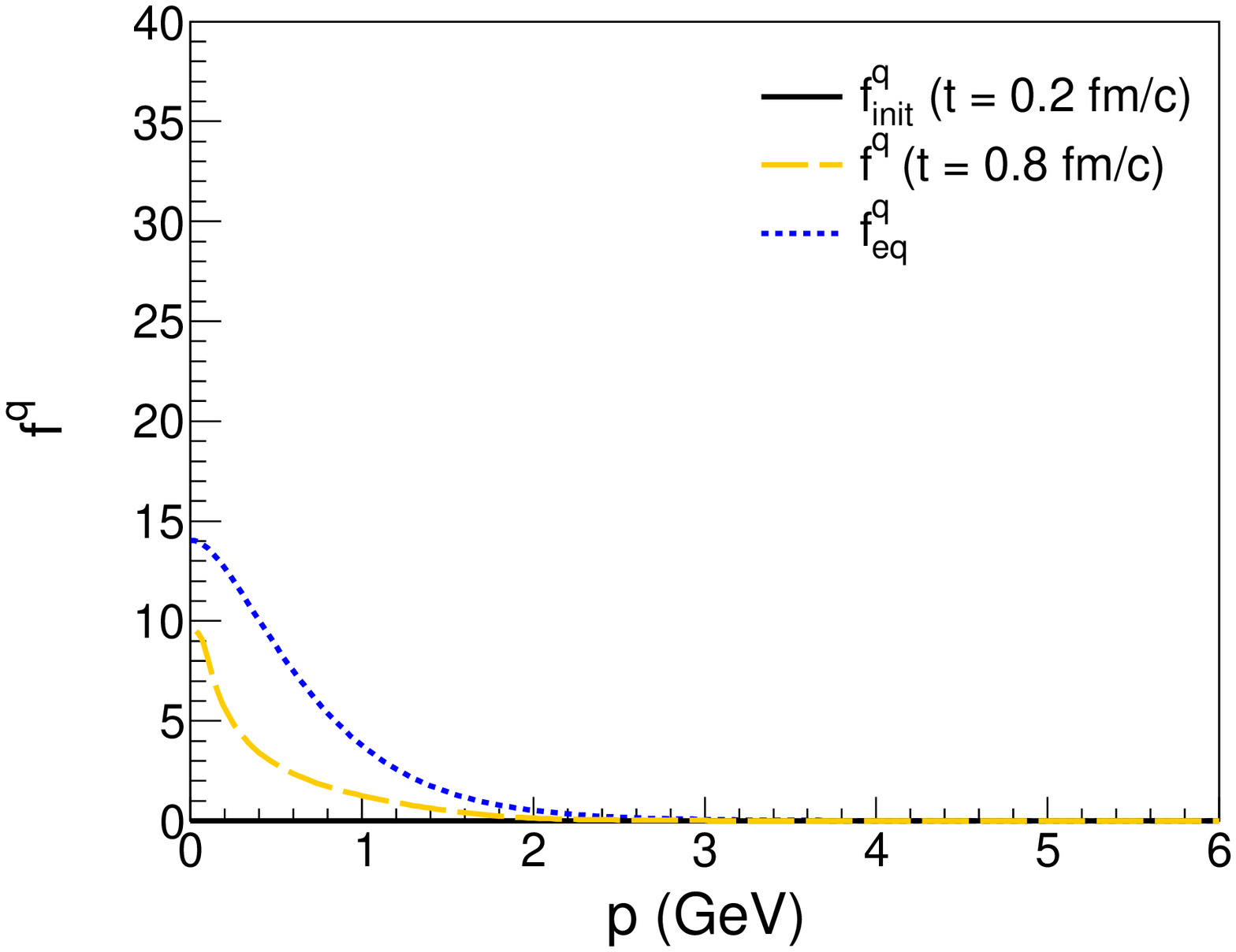}
\includegraphics[width=3.0in]{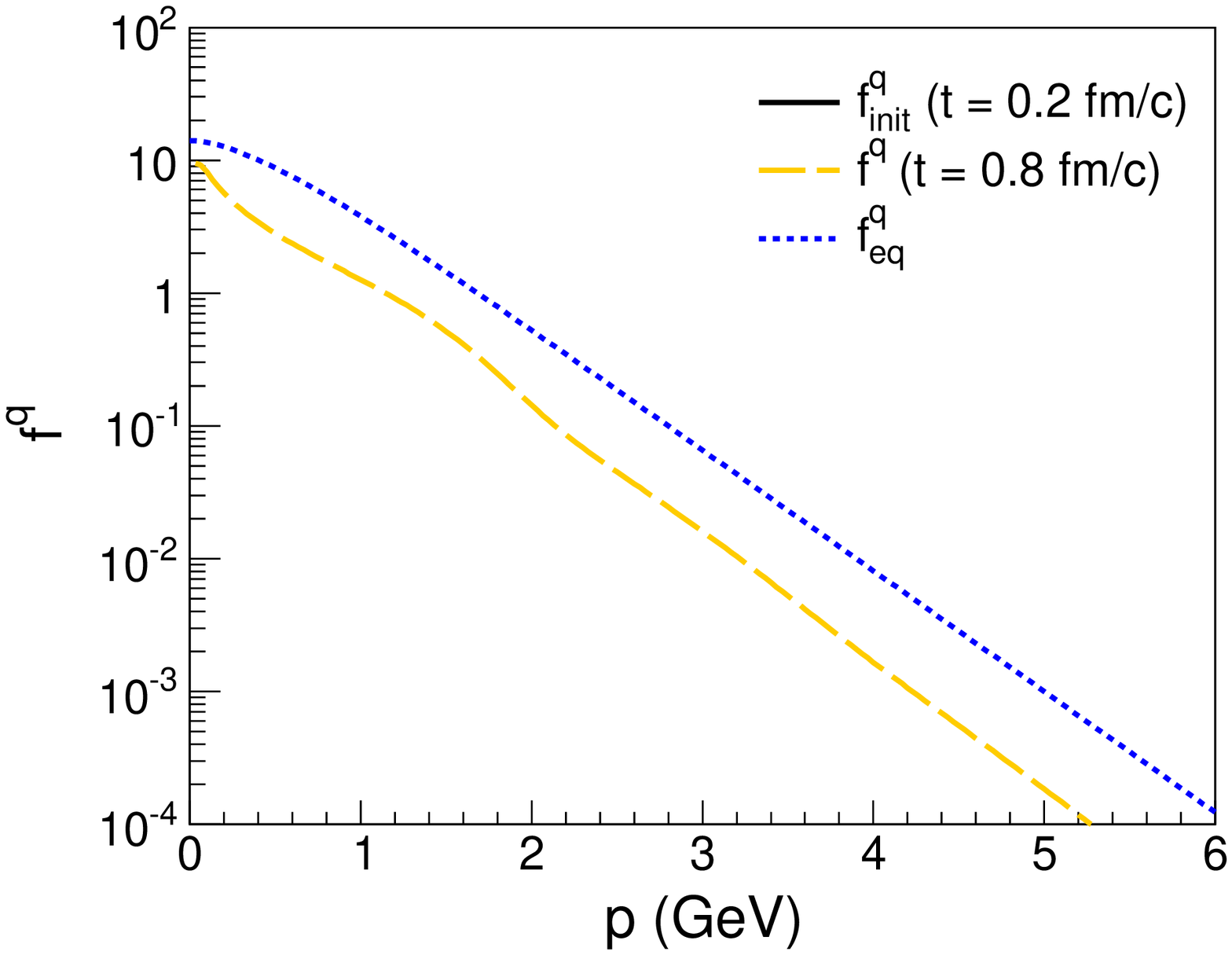}
\caption{\small (Color online) (a) The quark phase-space distribution in the initial state ($t=0.2$ fm, solid line) and during the time evolution ($t=0.8$ fm, dashed line) with $N_f = 3$ compared with that in equilibrium (dotted line). (b) The same figure on a logarithmic scale.
}
\label{fig:fq}
\end{center}
\end{figure}

The gluon and the quark distributions before and during the time evolution ($t = 0.8$ fm/$c$) are shown in Fig.~\ref{fig:fg} and \ref{fig:fq}, respectively. The number of flavors is set to $N_f = 3$. The initial conditions are the same as the ones in the previous calculation but the temperature of the equilibrium distributions is 0.458 GeV because of the additional degrees of freedom by quarks. One can see that the distributions again rapidly approach the equilibrium ones. Chemical equilibration of quarks and gluons are reasonably fast, but would be slower than gluon thermalization, possibly because of the difference in the coefficients for the splitting functions. The equilibration of the quark distribution occurs in $\sim$2 fm/$c$ in the current parameter settings. Also the shape of the quark distribution closely follows that of the gluon distribution since quarks are created via pair production from gluons. 

\begin{figure}[tb]
\begin{center}
\includegraphics[width=3.0in]{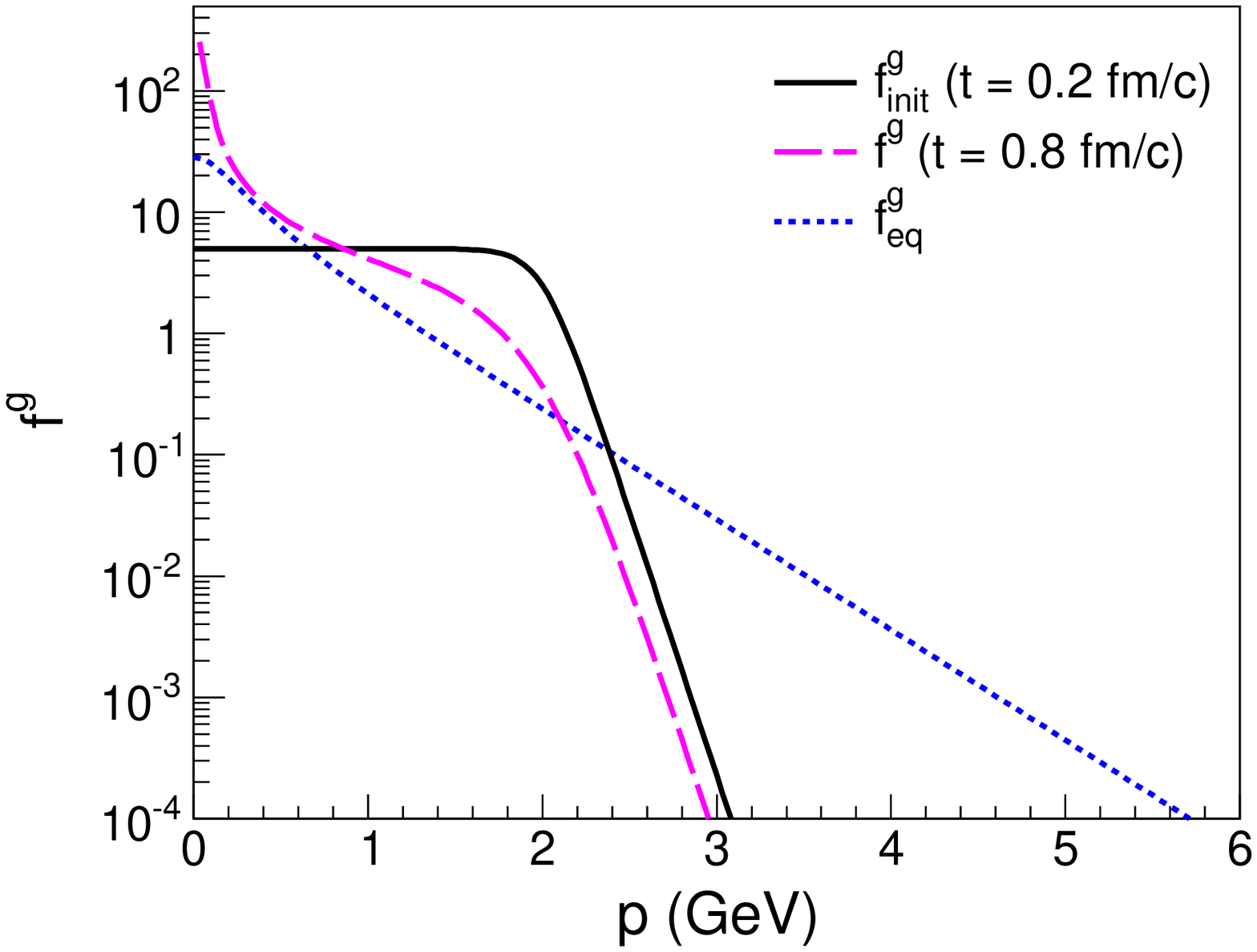}
\includegraphics[width=3.0in]{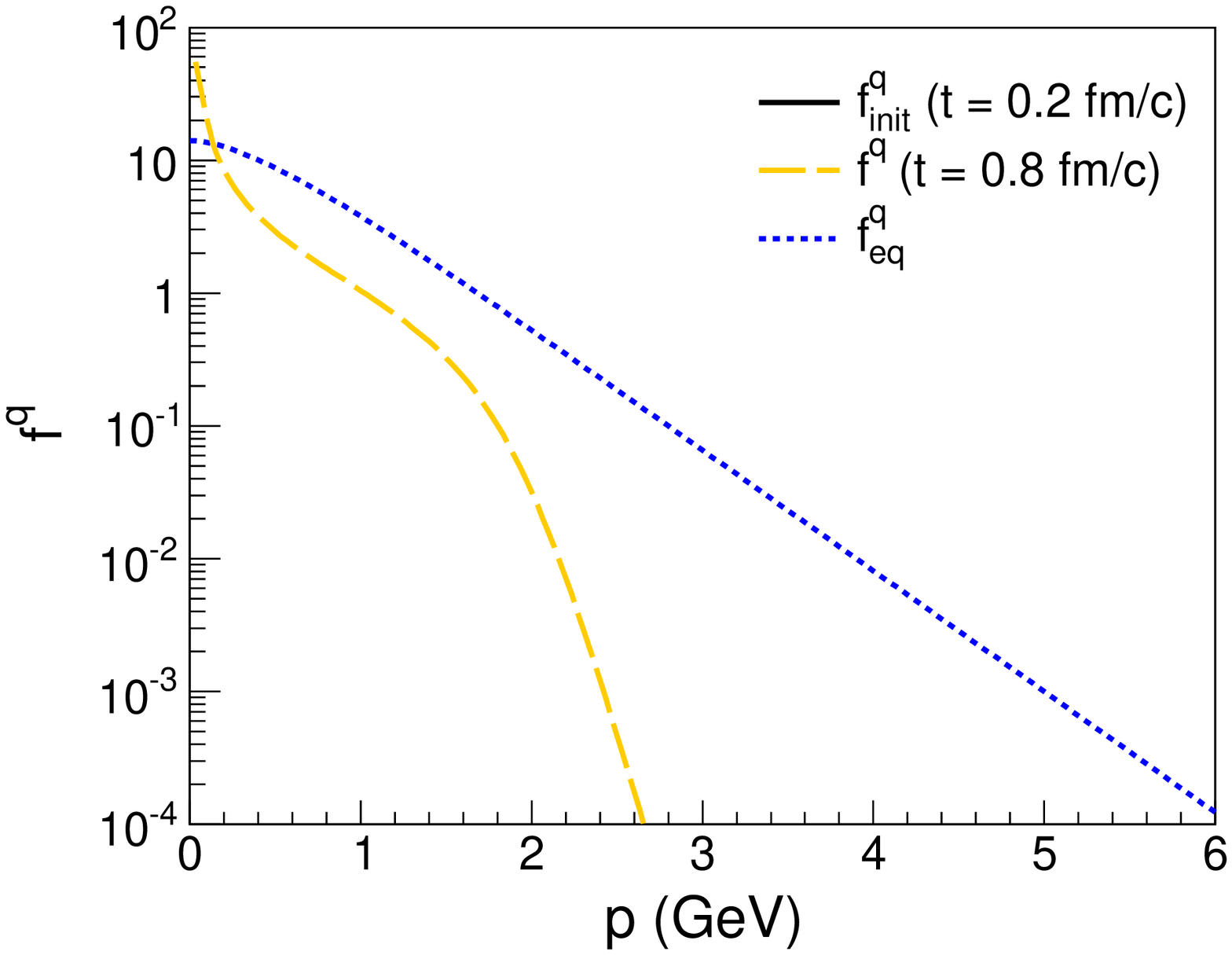}
\caption{\small (Color online) The phase-space distributions in the initial state ($t=0.2$ fm, solid line) and during the time evolution ($t=0.8$ fm, dashed line) compared with those in equilibrium (dotted line) with no recombination processes for (a) gluons and (b) quarks.
}
\label{fig:fsp}
\end{center}
\end{figure}

If the number of quarks are relatively small in comparison to the equilibrium one as suggested in Fig.~\ref{fig:fq}, no Fermi degeneracy pressure would be developed in the QGP. On the other hand, if there were no recombination processes or they were considerably weaker, the quark distribution can hit the limit of Pauli exclusion principle near $p=0$. Fig.~\ref{fig:fsp} shows the distributions after the time evolution with only splitting processes. This implies that recombination processes are the key in a quark-gluon system because when the splitting is dominant against the recombination near the vanishing momentum, the fermion production would be suppressed due to its quantum statistics and chemical equilibration could become significantly slower.

\section{Discussion and Conclusion}
\label{sec:4}
\vspace*{-2mm}

We have developed a model description of thermal and chemical equilibration of a gluon matter in high-energy heavy ion collisions based on collinear splitting of partons. The rapid production of low-momentum gluons from the collinear splitting processes encourages the early thermalization. Recombination is introduced so the system would be stable at the maximum entropy state. Effects of medium interaction are taken into account via relativistic Fokker-Planck equation, which encodes off-shell conditions in the coefficients $A^i$ and $B^{ij}$. 
It should be noted that the diffusion also encourages isotropization. 

Numerical demonstration in a one-dimensional non-expanding system shows that time evolution in the collinear splitting model with the decay rate and the momentum smearing coefficients implied from a weakly-coupled picture leads to thermalization of a color glass condensate-like gluon distribution in a very short time in a pure gauge system. This implies that the processes are essential in the early stages of the high-energy nuclear collisions. Chemical equilibration of quarks and gluons, as well as thermalization of quarks, are also reasonably fast given the weak coupling nature of the model, but could be slower than gluon thermalization. Note that the coefficients are roughly estimated and refinement is necessary for more quantitative discussion.

It would be important to numerically investigate the magnitude of each process more precisely in three-dimensional systems, especially because it is required for the discussion on isotropization. Effects of expansion in the heavy ion system can also be implemented as discussed in Appendix~\ref{sec:A}. Running coupling may be introduced for more accuracy. 
Further future prospects include the application of our model to the later stages in heavy ion collisions. The off-equilibrium corrections to the distribution might be attributed to the initial dissipative quantities for the fluid dynamic stage \cite{Israel:1979wp}, which is an interesting topic by itself because so far little is known about those quantities. 
It is also important that if chemical equilibration is not complete at the onset of the thermalized stage, one might need to have a two-component fluid for the description of the QGP dynamics. 

\begin{acknowledgments}
The authors would like to thank T.~Hatsuda, Y. Hatta, K.~Itakura, and R.~Venugopalan for valuable comments. The work is supported in part by RIKEN Special Postdoctoral Researcher program and a research grant from the U.S. Department of Energy (DE-FG02-05ER41367).
\end{acknowledgments}

\appendix

\section{EFFECTS OF EXPANSION}
\label{sec:A}
\vspace*{-2mm}

The effect of expansion in a heavy ion system can be included by introducing the drift term individually to the equations of motion for gluon and quark distributions: 
\begin{eqnarray}
\frac{\partial f}{\partial t}  + \frac{p_z}{E} \frac{\partial f}{\partial z} &=& \mathcal{C}_\mathrm{sp} + \mathcal{C}_\mathrm{rc} + \mathcal{C}_\mathrm{FP}.
\end{eqnarray}
The expansion term can be alternatively expressed as
\begin{eqnarray}
\frac{p_z}{E} \frac{\partial f}{\partial z} = - \frac{p_z}{t} \frac{\partial f}{\partial p_z}\end{eqnarray}
at $z=0$ for boost invariant expansion in $z$ direction \cite{Baym:1984np} since $\tau = t$ at mid-rapidity. The drift term can be interpreted as an external force decreasing with time that expands the system. In the absence of the splitting, the recombination and the Fokker-Planck terms, this equation allows a free-streaming solution $f(t, p_z) = f(t_0, p_z \frac{t}{t_0})$. 

The equation implies that the transverse momentum distribution would not be directly modified by expansion at $p_z = 0$. However the thermalization dynamics is indirectly affected since the decrease in the temperature alters the thermal equilibrium distribution itself.

\bibliography{basename of .bib file}

\end{document}